\UseRawInputEncoding
\documentclass[a4paper,10pt,oneside]{article}
\usepackage{icad2021,amsmath,epsfig,times,url}
\usepackage[pdftex,
            pdfauthor={Tim Ziemer & Navid Mirzayousef Jadid},
            pdftitle={Developing, Distributing and Marketing Sonification Apps: A Case Report},
            pdfsubject={preprint},
            pdfkeywords={Sonification Apps, Marketing, Promotion, Advertisement, Sonification Tools}]{hyperref}
\usepackage{pdfcomment}
\usepackage{multirow}
\usepackage{doi}
\usepackage[utf8]{inputenc} 

\title{RECOMMENDATIONS TO DEVELOP, DISTRIBUTE AND MARKET SONIFICATION APPS
}

\twoauthors{Tim Ziemer} {University of Bremen \\ Bremen Spatial Cognition Center \\ Bremen, Germany  \\ {\tt ziemer@uni-bremen.de}}
{Navid Mirzayousef Jadid} {University of Bremen \\ Computer Science Department \\ Bremen, Germany  \\ {\tt navid@uni-bremen.de}}

\begin{document}
\ninept
\maketitle

\begin{sloppy}

\begin{abstract}
After decades of research, sonification is still rarely adopted in consumer electronics, software and user interfaces. Outside the science and arts scenes the term \emph{sonification} seems not well known to the public. As a means of science communication, and in order to make software developers, producers of consumer electronics and end users aware of sonification, we developed, distributed, and promoted \emph{Tiltification}. This smartphone app utilizes sonification to inform users about the tilt angle of their phone, so that they can use it as a torpedo level. In this paper we report on our app development, distribution and promotion strategies and reflect on their success in making the app in particular, and sonification in general, better known to the public. Finally, we give recommendations on how to develop, distribute and market sonification apps. This article is dedicated to research institutions without commercial interests.
\end{abstract}

\section{Introduction}
\label{sec:intro}
Even though sonification has been proven to be useful for navigation \cite{matti,vadis,acta}, recognition of color intensity \cite{niklas}, sports performance \cite{nina}, patient monitoring \cite{intelligibility,schwarzziemer}, as an equivalent of scatterplots \cite{graphs} and many other purposes, it has not yet reached mainstream popularity \cite{killer,uisurvey}.


The idea of a \emph{killer application} is quite present in the sonification community. \cite{killer} describes the killer app as ``a novel use case that leads to ubiquitous adoption of sonification and thereby cements its relevance as a design method''. Similarly, \cite{oxford} describes it as being ``so convincing that it would make people `buy into' the idea of sonification in general, contributing to its acceptance''. Naturally, there is no need for a single killer app. Many researchers express the need for available sonification demonstrators and practical applications in general \cite[pp. 242ff]{scaletti} \cite[pp. 2f]{barrassthesis} and agree that ``Sonification will gain significant momentum once several specific applications become widely used.'' \cite[p. 24]{report}



With \emph{Tiltification} \cite{tilt} we aimed at contributing to the breakthrough of sonification. Tiltification is a smartphone app in which the two tilt angles of the smartphone are communicated through our two-dimensional psychoacoustic sonification \cite{ziemerschicad}. This way the sound serves as an auditory (bullseye) spirit level.

The development, distribution, and marketing of Tiltification was conducted during the \emph{Sonification Apps} master's project for Computer Science and Digital Media students at the University of Bremen from winter semester 2020/2021 to summer semester 2021.

In the paper at hand we report on the reception of Tiltification by the public and try to deduce which of our strategic decisions helped to popularize our app in particular, and sonification in general. 
Finally, we give recommendations on how to develop, distribute and market your own sonification apps to contribute in the widespread adoption and use of sonification.

\section{Our Strategy}
In 2020, one year prior to Tiltification, we has already developed and released a sonification app: the CURAT Sonification Game \cite{curat}. Our motivation for the CURAT Sonification Game was to evaluate our three-dimensional psychoacoustic sonification \cite{preprinto,icad2019} in a longitudinal and cross-sectional study from a population that is much larger than the $7$ \cite{cars} to $18$ \cite{jmui} subjects that we were able to recruit for experiments in the lab, through gamification.

Meanwhile, our motivation for Tiltification was science communication: We wanted to make sonification better known to the public by providing a useful app whose sonification would be a clear benefit over conventional, graphical solutions. As we had little prior experience, we made a number of strategic decisions that were supposed to serve our goal to reach as many people as possible with our app. These concerned the development, distribution channels and the marketing of the app.

\subsection{Development}
\label{development}
Our strategic decisions to make Tiltification successful can be summarized under $4$ terms:
\begin{enumerate}
    \item slim and easy to use
    \item accessible to visually impaired people
    \item multi-platform and multi-device
    \item multiple languages
\end{enumerate}


\subsubsection{Slim and Easy to use}
There is consensus in the sonification community that ``(\ldots) until there are intuitive, efficacious applications, skeptics will adhere to current display solutions'' \cite[.p 24]{report}. Thus, it is likely that if the new technique is too foreign from daily practice, people will not give it a chance \cite{oxford}. On the other hand, if the sonification app makes too many promises, clients may develop an ``expectation creep'' \cite[p. 142]{davidbuch}.

In order to be easy to use and not too different from conventional apps, we decided to create an app that serves as a simple and well-known tool: a two-dimensional (bullseye) spirit level. This way every user may easily understand what the app is for and what the sonification could potentially offer. The sonification of the two tilt angles is combined with a conventional visualization and text. Our idea was that this would help the user understand the sonification.

A spirit level indicated whether the device itself is leveled. If not, it signals in which direction(s), and, approximately, how far it needs to be tilted to get leveled. This is exactly what this app does, utilizing the two-dimensional sonification \cite{pomaziemerblacksch}, complemented by a conventional visualization with graphics and text.

Furthermore, we decided to provide a few utilities beyond the basic idea of the acoustic spirit level, but we made sure to keep them simple: the user can switch between one- and two-dimensional spirit level, between portrait and landscape mode, set a tilt-offset, and decide either to mute the sound, to play it only when the app is in the foreground, or to keep playing while the app is active in the background.

Lastly, we tried to keep Tiltification handy by a small file size. CURAT had a file size of $71.2$ MB, while we managed to keep Tiltification to a small $23.9$ MB on Android and $23$ MB on iOS.

\subsubsection{Accessibility}
To be accessible to visually impaired people, the app starts with its main functionality right away, without the need to set it up. This way even a completely blind person can use it. To be useful for visually impaired people, only a low number of large, iconic buttons with high contrast are being used. They are linearly aligned so that they can be reached via thumb, with a flat menu hierarchy. Little text is used, in a clear font without serifs, avoiding italics and all-caps, while using a large font size and a clear structure with large, bold headlines, short passages and short terms. This also improves the app's accessibility for people with dyslexia.

\subsubsection{Multi-platform and multi-device}
We developed the app in Flutter \cite{Payne2019}, which allows programming of cross-platform  apps with only one codebase, and which is compatible with libpd \cite{brinkmann_embedding_2011} to realize procedural audio, i.e., interactive sonification, rendered in Pure Data \cite[p. 152]{davidbuch}. The app was made compatible with smartphones and tablets with iOS or Android operating systems. It was tested and debugged on multiple smartphones and tablets.

\subsubsection{Multiple languages}
We decided to make the app available in English, as it is globally accepted \cite[p. 303]{marketing}, as well as in 
German, Spanish and Chinese. All in-app texts and app store listings, for both the Apple App Store and the Google Play Store, were translated by our team.

\subsection{Distribution}
\label{release}
To distribute the app we decided to host the APK for Android systems on our university server and create a listing on the two mainsteam app stores, the \href{https://apps.apple.com/de/app/tiltification/id1557133147}{Apple App Store} and the \href{https://play.google.com/store/apps/details?id=de.uni_bremen.informatik.sonification_apps}{Google Play Store}.




Developer Accounts for Apple and Google Play are free for governmental organizations. Apps are reviewed prior to publication, and they have to fulfill many guidelines. Furthermore, apps published in these stores, even free apps without advertisements or in-app purchases, are considered as U.S.-exports. Thus they are subject to U.S. Encryption and Export Administration Regulation \cite{cia}. As soon as you use encryption, be it a web link via \emph{https} or saving presets via \emph{AES} encryption, you need to submit a self-classification report to the U.S. Bureau of Industry and Security once or annually, depending on the classification of the app.


\subsection{Marketing}
\label{marketing}
Our marketing strategy was hybrid, including social media marketing, outreach to scientific communities, and a press release by the university, accompanied by an informative project website.

\subsubsection{Social Media Marketing}
\label{socialmedia}
Social media marketing is recommended to advertise apps \cite[pp. 207 \& 218ff]{macapps}\cite{500}\cite[pp. 115ff]{sandberg}.
Already during the conceptualization and development, we created an \href{https://www.instagram.com/tiltification/}{Instagram} and a \href{https://www.facebook.com/tiltification/}{Facebook} account for social media marketing. Through these channels we communicated our project goals, design decisions, called for beta testers, promoted the CURAT Sonification Game and generally sent out greetings with our logo, following some social media marketing tips \cite{500}.

We did not create a YouTube account dedicated to our app, but we produced \href{https://www.youtube.com/watch?v=SAQF2lO8zM8&list=PLVv3BMS8IIXEm748HgrkAvFtymaXLAHX6}{YouTube videos} that were collected within a playlist and released on the first author's YouTube account, as suggested as an advertisement and a service for potential users \cite[pp. 106f]{sandberg}.

\subsubsection{University Press Release}
\label{pressrelease}
Aiming to reach journalists we wrote a press release together with the university's communication and marketing staff. Press releases are a suggested app marketing strategy \cite[pp. 204 \& 236ff]{macapps}\cite[pp. 118f]{sandberg}, ideally written in both your national language and in English \cite[pp. 401ff]{marketing}. Our university press releases have around $400$ subscribers. However, the University of Bremen is a member of the \href{https://idw-online.de/en/}{idw scientific information service} who forward our press releases to roughly $18,000$ subscribers, including $4,200$ journalists, $360$ of them accredited.


\subsubsection{Scientific Communities}
\label{icad}
To reach people from audio-related scientific communities, we announced the release of our apps in scientific mailing list. E-mail marketing is an established strategy to advertise apps \cite[pp. 211ff]{macapps}.

Conference talks have even been suggested as one guerrilla marketing step for iOS and Mac apps \cite[p. 221]{macapps}



\subsubsection{Website}
\label{website}
We created a \href{https://sonification.uni-bremen.de}{Tiltification website} on a subdomain of our university website to provide interested people with reliable information on our app and our team. The website also features download links, an FAQ and an e-mail address, as suggested in \cite[p. 406]{marketing} and \cite[p. 113]{sandberg}.

\section{Method}
The goal of Tiltification was to bring as many people as possible in contact with our sonification app. The success is derived quantitatively by the number of downloads and active users, and qualitatively by the feedback that we received from the individuals and organizations.

The objective of this paper is to analyze the effectiveness of our strategic means to achieve our goal.

First, we examined the statistics that are provided by the Apple App Store and the Google Play Store and associated them with marketing events for both CURAT and Tiltification. Learning from analysis of app store statistics is recommended for app developers \cite[pp. 132ff]{sandberg}. These statistics include the time series of downloads, downloads per country, downloads per device type, downloads per source (referrer) and the number of active users per month.

In addition, we analyzed online responses through social media, forum discussions, inquiries via e-mail, reviews articles and reviews in the app stores.

Note that we cannot prove causal relationships between single events and download statistics, as
\begin{itemize}
    \item events may interfere,
    \item events may take time to take effect
    \item many latent variables may exist, 
    \item we have limited insight into printed media,
    \item we have no insight into private communication channels.
\end{itemize}
\section{Results and Discussion}
To understand which of our strategic decisions were meaningful in order to popularize Tiltification, we tried to link the decisions to a) the statistics provided by the app stores and b) online feedback.

\subsection{App Store Statistics}
App downloads per country are listed in Table \ref{table:countries} for Tiltification and the CURAT Sonification Game. The table only lists downloads from within the app stores, not the APK downloads. One can clearly see that Tiltification has more than $10$ times as many downloads as CURAT. Tiltification was downloaded $1.37$ times as often from the Google Play Store than from the Apple App Store. Both apps were mostly downloaded from Germany, but CURAT has a distinctly more international user base.
\begin{table}[ht]
\centering
\begin{tabular}{c | c c c }
 & (T) Apple & (T) Play & (C) Play\\
 \hline
All & $7,410$ & $10,155$ & $1,627$\\
Germany & $89.7$\% & $90.22$\% & $67.91$\%\\ 
Austria & $2.82$\% & $3.21$\% & $2.24$\%\\ 
Switzerland & $3.1$\% & $2.14$\% & $1.49$\%\\ 
Italy & $0.16$\% & $0.45$\% & \textit{NAN}\\ 
Spain & $0.18$\% & $0.47$\% & $3.73$\%\\ 
France & $0.16$\% & $0.3$\% & $5.22$\%\\ 
China & \textit{NAN} & $0.03$\% & \textit{NAN}\\ 
S. Korea & \textit{NAN} & $0.05$\% & \textit{NAN}\\ 
\end{tabular}
\caption{Number of downloads per country for Tiltification (T) in Apple App Store (Apple) and Google Play Store (Play), and downloads of the CURAT Sonification Game (C) in the Google Play Store.}
\label{table:countries}
\end{table}

Figure \ref{pic:pdownloads} shows the time series of Tiltification downloads from the Google Play Store. Posts from our social media accounts, the project day at the University (an event where computer science students present their project work to each other) and the presentation of our app at the ICAD conference had no visible impact. It was the \href{https://www.uni-bremen.de/en/university/university-communication-and-marketing/press-releases/detail-view/so-hoert-man-ob-das-bild-schief-haengt}{press release} that triggered an avalanche of downloads. The press release was accompanied by $2$ online articles from consumer electronics magazines (\href{https://www.heise.de/news/Tiltification-Akustische-Wasserwaage-fuers-Smartphone-6193037.html}{heise} and \href{https://www.iphone-ticker.de/tiltification-app-fungiert-als-akustische-wasserwaage-an-179736/}{iphoneticker}) and their social media posts that referred to their articles. The next day, $8$ additional online magazines from the field of science communication published articles on our app\footnote{Namely \href{https://hv.hansevalley.de/p/hansescience.html?m=1}{HanseValley},  \href{https://marketresearchtelecast.com/tiltification-acoustic-spirit-level-for-the-smartphone/156786/}{Market Research Telecast}, \href{https://www.mdr.de/wissen/faszination-technik/akustische-wasserwaage-per-app-uni-bremen-100_box-3846508809229059499_zc-3435bf4b.html}{mdr Wissen}, \href{https://www.pressetext.com/news/-tiltification-macht-handy-zur-wasserwaage.html}{pressetext}, \href{https://www.rappelsnut.de/tiltification-die-akustische-wasserwaage/}{Rappelsnut}, \href{https://thescienceplus.com/news/newsview.php?ncode=1065615399925491}{theSCIENCEplus}, \href{https://www.scinexx.de/news/technik/handy-app-als-akustische-wasserwaage/}{scinexx}, \href{https://www.vdi-nachrichten.com/technik/forschung/mit-der-wasserwaage-hoeren-ob-alles-im-lot-ist/}{vdi nachrichten}.}. During these two days $8$ tweets on Twitter referred to the press release or either of the articles. Within $6$ days the Google Play Store counted $7,329$ downloads, mostly from Germany. Such a \emph{peak period} is common for Android apps \cite[p. 137]{sandberg}. A second, smaller peak $2$ weeks later can be observed and may be associated with the release of $2$ additional online articles and posts in $3$ scientific mailing lists. After the main hype and its straggler, no online or printed magazine article had a visible impact on the download numbers.

\begin{figure}[!ht]
\centering
\includegraphics[width=8cm]{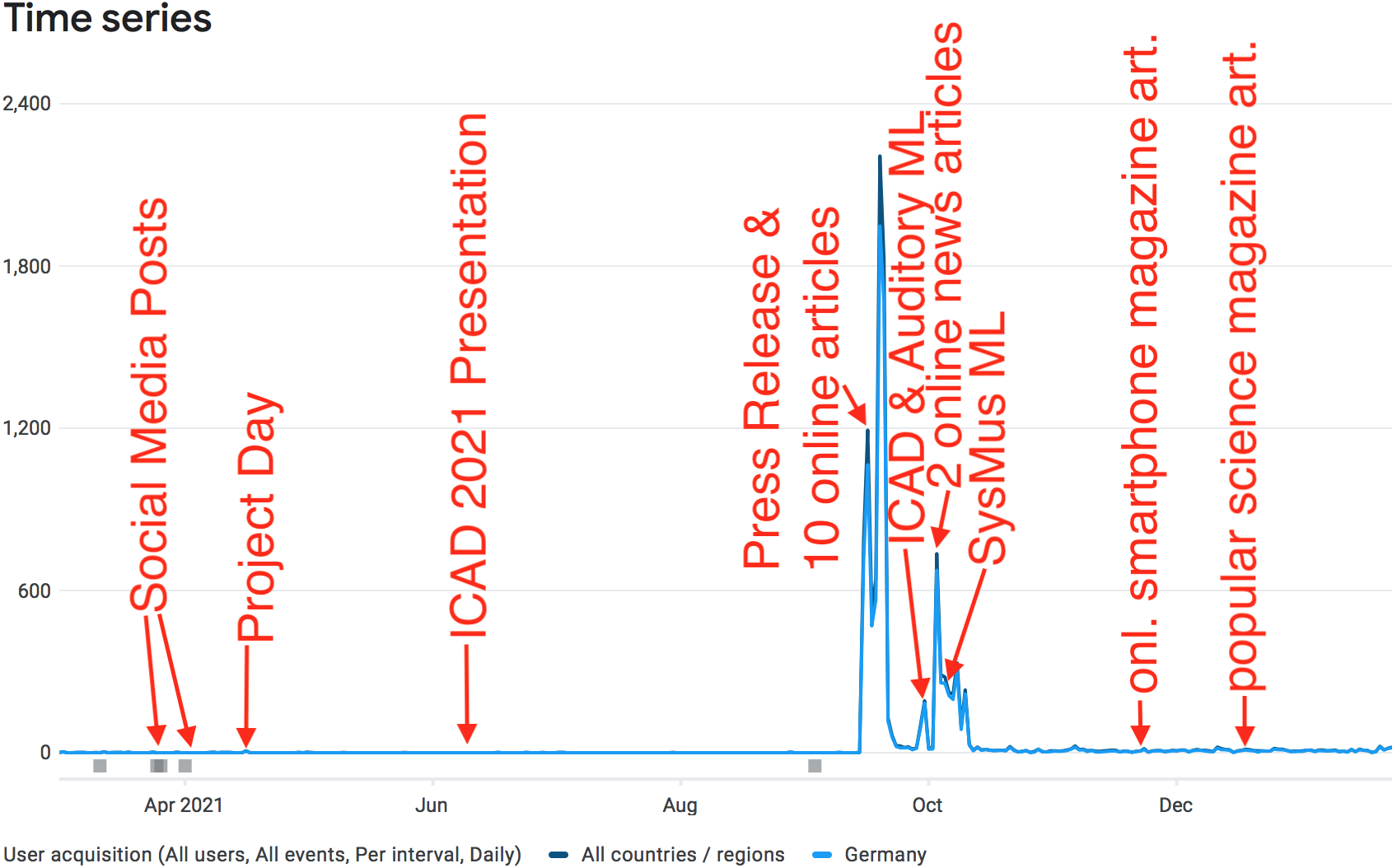}
\caption{Daily downloads of Tiltification from the Google Play Store. Within one day after the press release $10$ online magazines published articles about Tiltification, which led to $813$ and $1,229$ downloads on these days and $5,287$ during the following $4$ days. Posts in mailing lists had a visible effect, social media posts, the ICAD presentation and some online articles had no direct impact.}
\label{pic:pdownloads}
\end{figure}

Figure \ref{pic:curatdls} shows the time series of CURAT downloads. Here, practically all peaks can be associated with a marketing event. Each post in a scientific mailing list produced a peak. The press release did not create a peak itself, but led to a public radio report and an article in a computer magazine, which produced peaks. 

It is difficult to identify why the Tiltification press release led to $20$ online articles, $1$ magazine article and almost $20,000$ downloads, whereas the CURAT press release only led to $4$ online articles, $2$ magazine articles and $1$ radio report and some hundred app downloads. We identified $4$ differences between the CURAT and the Tiltification press release that may have contributed to the difference in spread and reception:
\begin{enumerate}
    \item The CURAT press release was dedicated to research in the surgical field, Tiltification targeted general smartphone users
    \item CURAT was communicated as an ongoing research project, Tiltification, as a  product
    \item CURAT was supposed to help the creators, whereas Tiltification was an offer for the users
    \item CURAT is a game, Tiltification a handy tool
\end{enumerate}
It seems to be wise to dedicate the press release to a broad audience, communicate it as a product and an offer for the user. Furthermore, utilities have fewer competitors than games \cite{statistica}.



Specialized articles dedicated to a narrow target group, like the university's online magazine for students and intellectuals, the ICAD community spotlight and a hospital magazine had no impact on the number of downloads. The same holds true for the presentation at the ICAD conference. The last peak was certainly owed to the fact that CURAT was mentioned and linked in the Tiltification press release.

\begin{figure}[!ht]
\centering
\includegraphics[width=7.5cm]{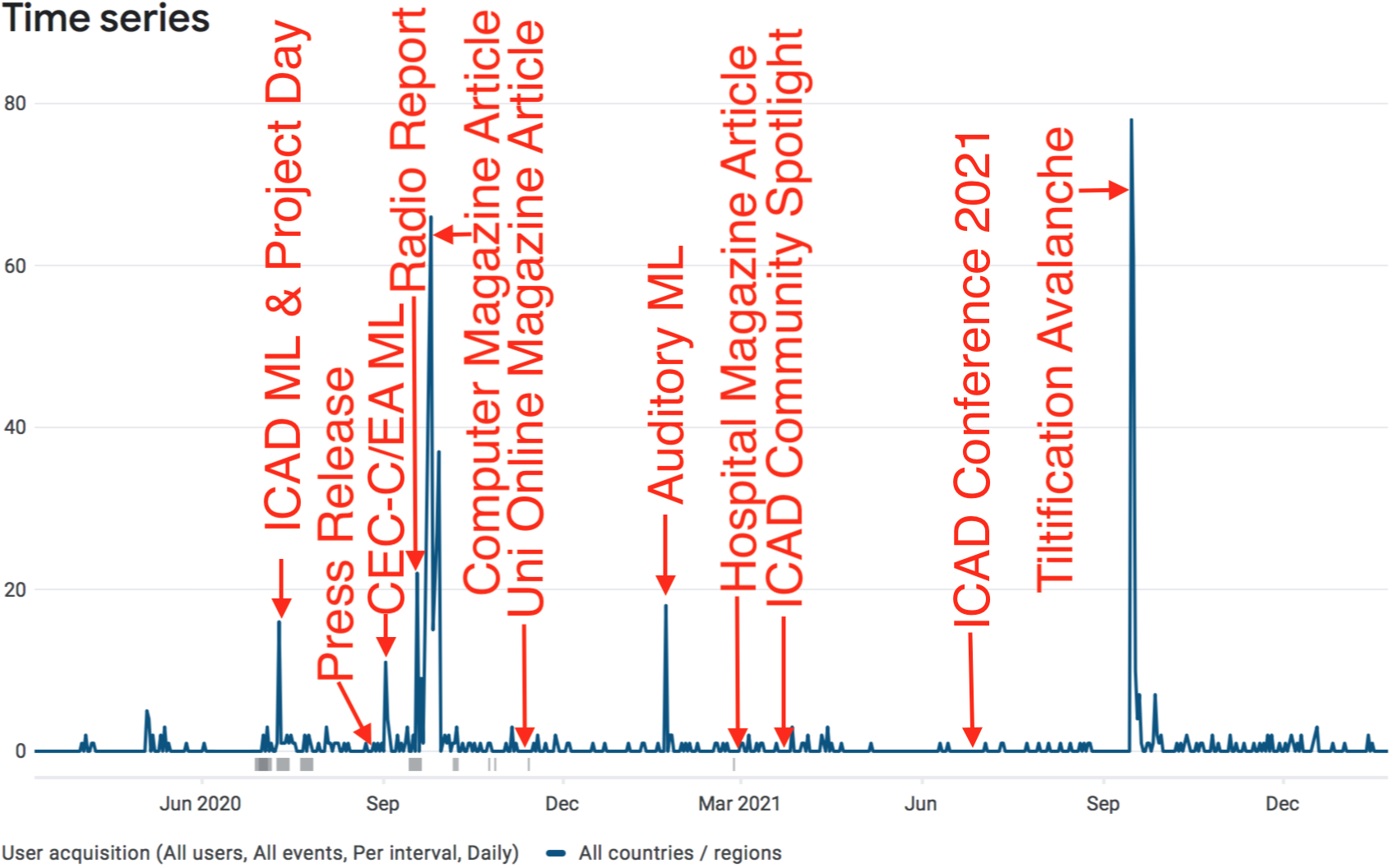}
\caption{Daily downloads of the CURAT sonification game. Many events directly caused a small but distinct download peak, like the posts in audio research mailing lists (ML), and the radio and magazine articles and press release that were dedicated to a broad audience. The conference presentation and articles for narrow target groups (like the medically-centered press release, the hospital magazine and the university's online magazine) did not produce peaks themselves.}
\label{pic:curatdls}
\end{figure}

Figure \ref{pic:idownloads} shows the time series of Tiltification downloads from the Apple App Store. The general course is identical with the plot from the Google Play Store downloads, with one large peak starting the day of the press release, and a second, smaller peak two weeks later. Only details of the peaks look different.

\begin{figure}[!ht]
\centering
\includegraphics[width=8cm]{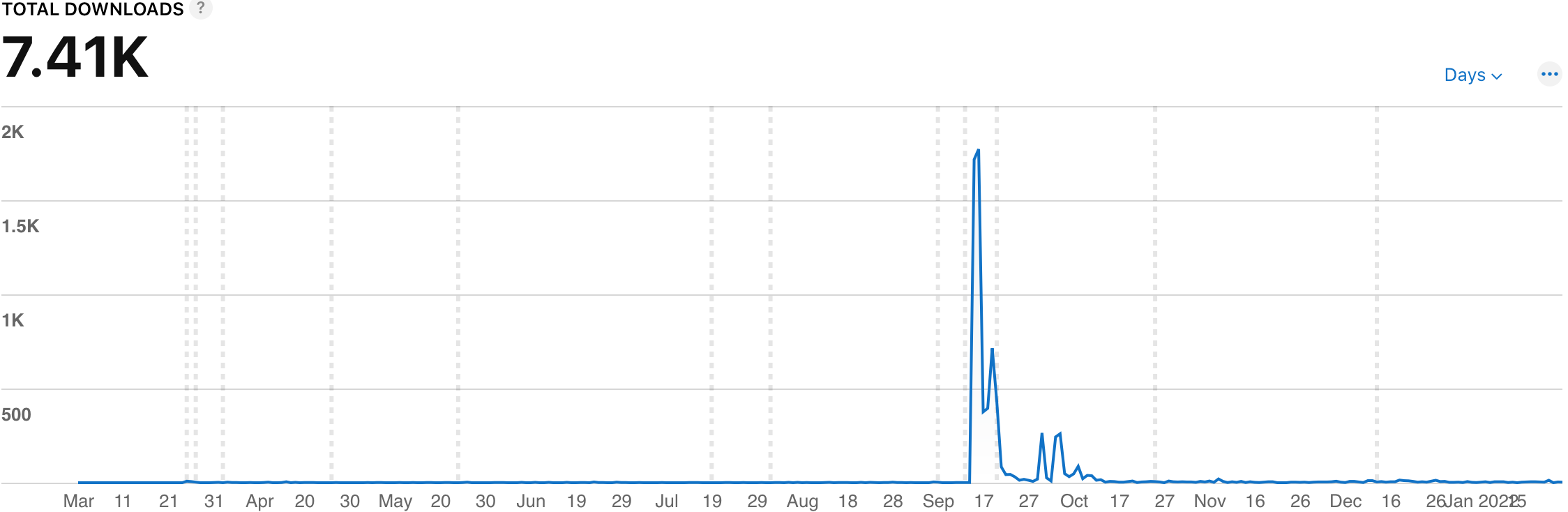}
\caption{Daily downloads of Tiltification from the Apple App Store. The plot bears strong similarity to the time series from the Google Play Store.}
\label{pic:idownloads}
\end{figure}

Table \ref{table:type} lists the percentage of Titification downloads per device type. The vast majority of downloads was for smartphones.

\begin{table}[ht]
\centering
\begin{tabular}{c | c c c}
 & (T) iOS & (T) Android & (C) Android\\
 \hline
Smartphone & $89$\% & $96.8$\% & $92.7$\%\\
Tablet & $10.8$\% & $2.9$\% & $6.5$\%\\
\end{tabular}
\caption{Percentage of downloads for smartphone and tablet. The remaining percentage is listed as ``Desktop'' or ``Unreported'' in the respective stores.}
\label{table:type}
\end{table}

Table \ref{table:source} shows how users were referred to the Tiltification app. Most users found the app via search function in the respective stores, followed by referrals from websites and apps. Note that these numbers may be biased: Most online articles did to link to the app store listings, so readers had to search for the app in the respective stores. This assumption is supported by the fact the most frequent search term of Tiltification downloaders in the Google Play Store was ``Tiltification'' ($69.1$\%). The Apple App Store provides additional detail: $67.8$\% of the web referrals were from uni-bremen.de, $29$\% from iphone-ticker.de (whose article on Tiltification did in fact link to the Apple App Store), $1.8$\% from Google and $1.3$\% computerwissen.de (interestingly, we were not able to find a single reference to Tiltification on computerwissen.de).

\begin{table}[ht]
\centering
\begin{tabular}{c | c c }
 & iOS & Android \\
 \hline
App Store Search & $63.9$\% & $53.1$\% \\
Web referral & $22$\% & \multirow{2}{*}{$\sum 46.2$\%}\\
App Referral & $10.2$\% & \\
App Store Browse & $3.7$\% & $0.7$\%\\
\end{tabular}
\caption{Tiltification download sources and their overall shares.}
\label{table:source}
\end{table}



Figure \ref{pic:mau} shows the number of monthly active Tiltification users on Android. After the peak with $10,071$ monthly users, a user base of over $1,500$ monthly users had established, though slowly decreasing. The app is still installed on over $6,000$ android devices. As many users certainly need their spirit level less frequently than once per month, the number of occasional users may lie between $1,500$ and $6,000$.

\begin{figure}[!ht]
\centering
\includegraphics[width=8cm]{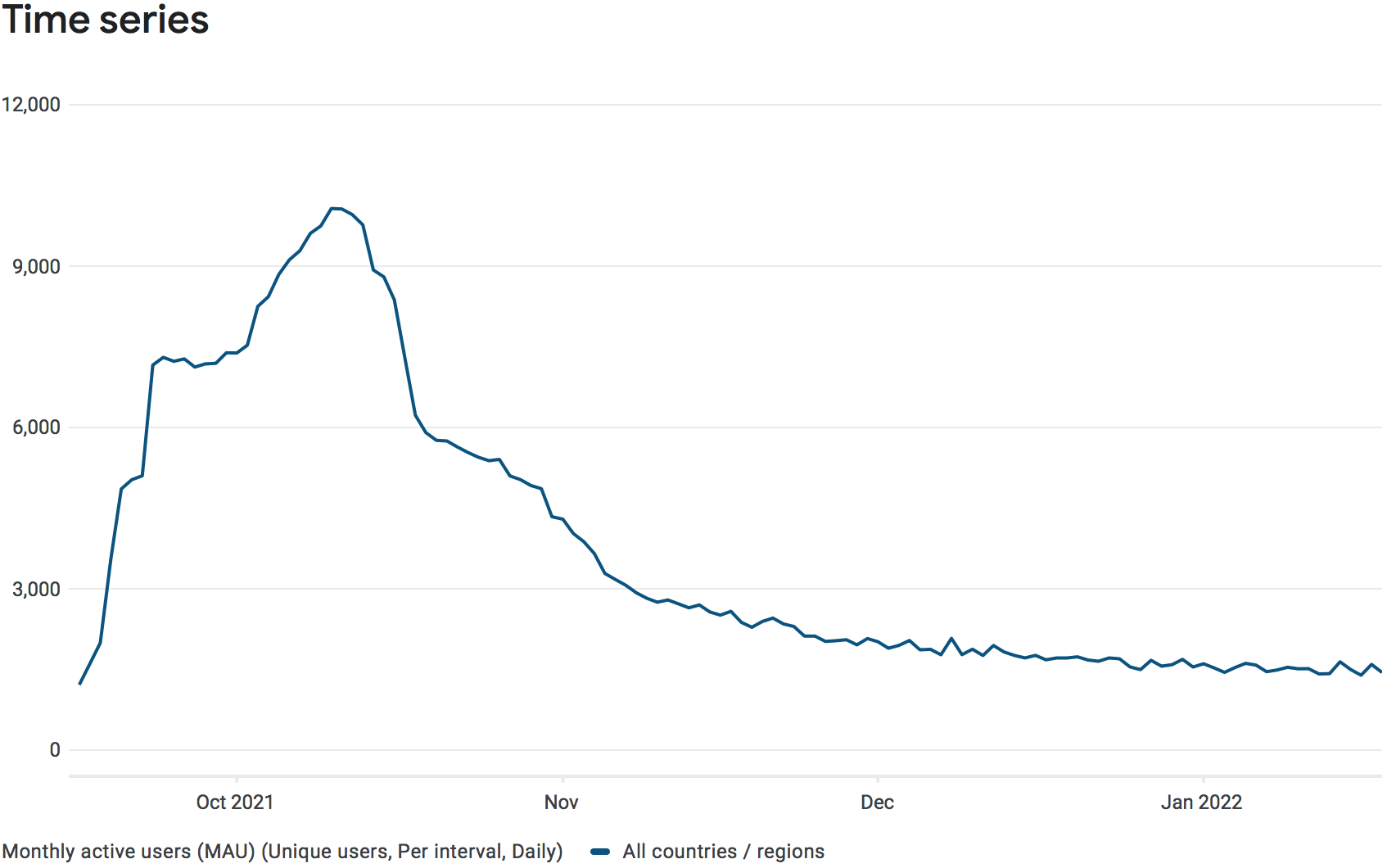}
\caption{Monthly active Tiltification users on Android. After a short hype with over $10,000$ users, we observed a transition from $1,900$ users in early December to $1,600$ users in early January.}
\label{pic:mau}
\end{figure}

Figure \ref{pic:curatdevices} shows how many Android devices had CURAT installed since its release. Clearly, the promotion through scientific mailing lists acquired an international user base, whereas the Tiltification press release mostly invoked German users.

\begin{figure}[!ht]
\centering
\includegraphics[width=7.5cm]{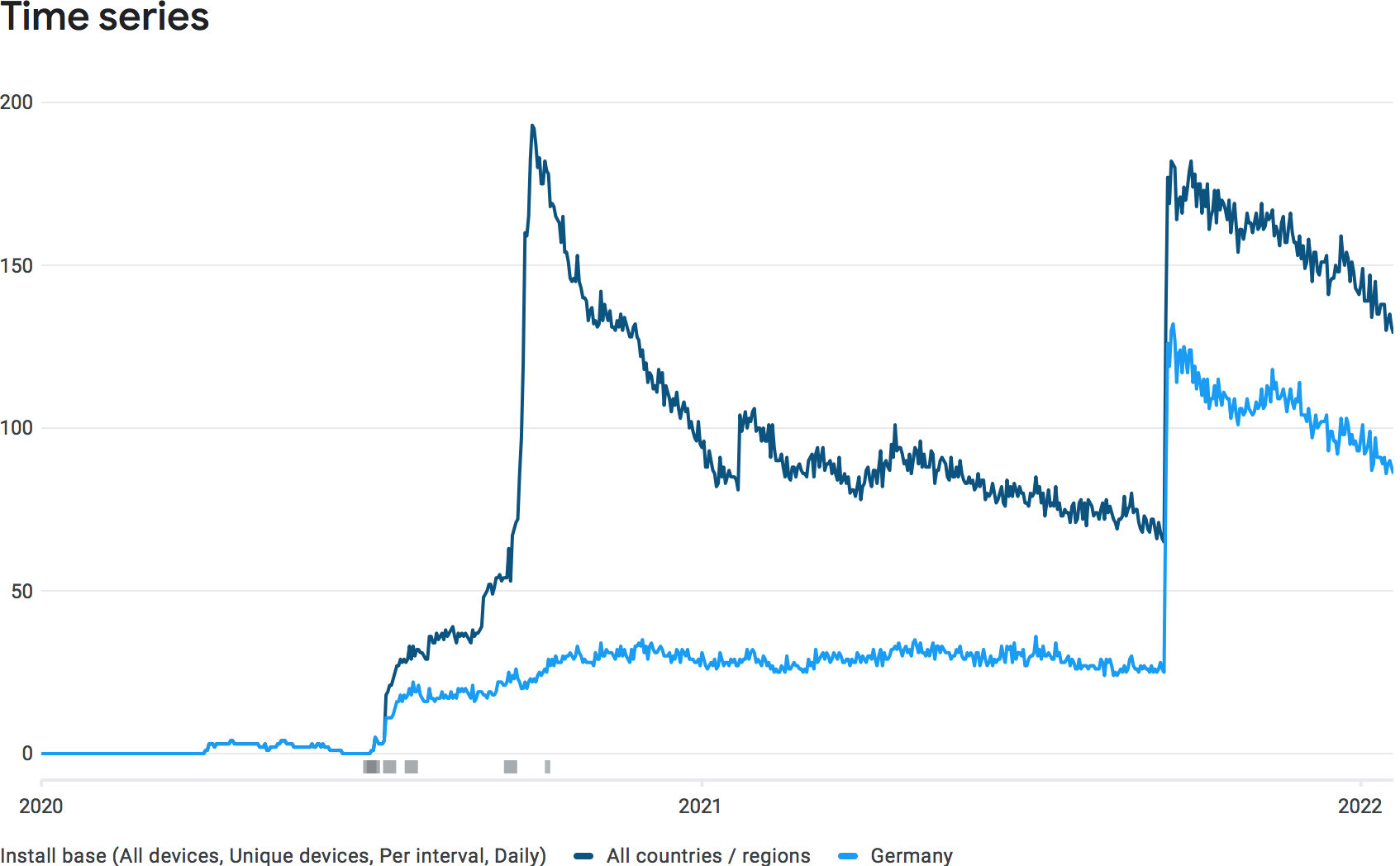}
\caption{Time series of install base, i.e., Android devices that had CURAT installed. The first wave was quite international, the second wave, initiated by the Tiltification press release, was dominated by German users.}
\label{pic:curatdevices}
\end{figure}

\subsection{Online Responses}
\textbf{Distribution:} Our APK was redistributed by the $4$ APK providers \href{https://apkpure.com/de/tiltification/de.uni_bremen.informatik.sonification_apps}{apkpure},
\href{https://apkpremier.com/details/de-uni_bremen-informatik-sonification_apps}{apkpremier},  \href{https://apk-dl.com/tiltification/de.uni_bremen.informatik.sonification_apps}{apk-dl} and \href{https://steprimo.com/android/us/app/de.uni_bremen.informatik.sonification_apps/Tiltification/}{ste primo}.

Tiltification was listed as a useful app for visually impaired and blind people by the \href{http://www.bfw-wuerzburg.de/download/appliste-iphone.pdf}{educational center for blind and visually impaired people Würzburg} and as a suitable app for the everyday life of chemists by \href{https://analytik.news/links/laborsoftware/smartphone-apps.html}{Analytik News} and is  
rank $593$ (of $1,574$) in  \href{https://www.chip.de/download/59063_Tools-Dienste/wochen-charts/}{CHIP's weekly top download ranking} (Week January 24 to 30 2022).

\textbf{App Ratings and Reviews:} Table \ref{table:ratings} lists the user ratings from different platforms, which generally positive. Reviews from the App stores ranged from positive to negative: Some users liked how informative the sound and the visualization were, suggested features and improvements, partly stated problems interpreting the sound and graphics, and partly criticized the sound as being unpleasant or annoying. One person complained that the sound kept running in the background (which is a feature that can (de-)activate inside the app), one user reported crashes and another user even gave an explanation for those users who were unable to interpret the sound. One user changed his $2$ star-rating to $5$ stars after we've explained the sound options in a reply.
\begin{table}[ht]
\centering
\begin{tabular}{c | c c }
 & users & mean rating \\
 \hline
\href{https://apps.apple.com/de/app/tiltification/id1557133147}{Apple App Store} & $19$ & $4.6/5$ \\
\href{https://play.google.com/store/apps/details?id=de.uni_bremen.informatik.sonification_apps}{Google Play Store} & $42$ & $4.429/5$ \\ 
\href{https://www.chip.de/downloads/Tiltification-Wasserwaage-iPhone-_-iPad-App_183818739.html}{CHIP iOS} & $20$ & $4.4/5$ \\ 
\href{https://www.chip.de/downloads/Tiltification-Wasserwaage-Android-App_183818812.html}{CHIP Android} & $123$ & $3.9/5$ \\ 
\end{tabular}
\caption{Number of ratings and average rating of Tiltification.}
\label{table:ratings}
\end{table}

\textbf{Review Articles:} All $3$ review articles were positive:  \href{https://www.chip.de/news/Intelligente-Werkzeug-App-Studenten-machen-es-besser-als-Apple_183818881.html}{CHIP} stated that the apps was easy to use, more precise than a physical spirit level and the spirit level in Apple's pre-installed Measure app. They praised the acoustic feedback and their overall rating was ``good''. \href{https://smartphonemag.de/apps/die-besten-neuen-apps-vom-november-2021/}{Smarthone} stated that the app was useful assistant for everyday life and the overall rating was ``good'', and their printed magazine 1/22 listed Tiltification as one of the best new apps (``Die besten neuen Apps'', p. 46).  \href{https://www.mdr.de/wissen/faszination-technik/akustische-wasserwaage-per-app-uni-bremen-100_box-3846508809229059499_zc-3435bf4b.html}{mdr Wissen} called Tiltification ``recht intuitiv'' (quite intuitive) and praised that the app would enable users to do home improvement on their own.


\textbf{Social Media:} Our Instagram account followed $229$ institutions and had $63$ followers. On average, our $24$ were liked $4$ times ($0$ to $15$), one post was commented once, and our videos were watched $35.9$ times ($20$-$48$). On Facebook we had $14$ followers. On average, our $17$ posts were liked $0.94$ times and none was commented.

These numbers highlight that creating social media accounts dedicated to a single sonification app project only reached a very small audience.

We found $12$ Tweets about Tiltification. On average, they received $2.8$ likes and were retweeted $1.2$ times, while none was commented. All tweets referred to the press release or one of the online magazine articles. The tweet by the registered association of blind and visually impaired people in Hessen, Germany (``Blinden- und Sehbehindertenbund in Hessen e.V.'') appreciated that Tiltification was accessible for visually impaired people, which was retweeted by the association of blind and visually impaired people in Austria (``Hilfsgemeinschaft der Blinden und Sehschwachen Österreichs''). 

On average our $11$ Tiltification YouTube videos had $206.5$ views ($63$ to $1,271$) and received $3.1$ thumbs up ($0$ to $23$). Two videos were commented once. Our $3$ tutorial videos were watched $133$ times on average ($118$ to $149$), whereas our $4$ demo videos were watched $67.5$ times ($63$ to $77$). Our teaser video was watched $1,271$ times. The teaser video was embedded in the Google Play Store listing. We referred to the tutorial videos as response to a review in the Google Play Store. It seems that it was the Play Store listing that produced YouTube views, rather then the other way around. This is underlined by our observations on the CURAT videos. The English CURAT teaser video embedded in the Google Play Store listing had $539$ views, whereas the German teaser counted $141$ views, and the video presentation for the project day (the day at which computer science and digital media students present their project results to each other) counted $104$ views. Consequently, YouTube videos seem to serve as a service for existing users rather than a promotion to acquire new users.

\textbf{Online forums:} Three online articles had comment sections that were used by $2$ (\href{https://www.rappelsnut.de/tiltification-die-akustische-wasserwaage/}{rappelsnut}), $27$ (\href{https://www.iphone-ticker.de/tiltification-app-fungiert-als-akustische-wasserwaage-an-179736/}{iphoneticker}) and $34$ (\href{https://www.heise.de/news/Tiltification-Akustische-Wasserwaage-fuers-Smartphone-6193037.html}{heise}) readers. Comments concerned the respective article, the app and some off-topics. Feedback on the app ranged from praising to dismissive. Some readers found the sonification innovative and useful. This is what we were hoping for. Some readers appreciated that the app was \emph{truly free}, i.e., without advertisements, in-app purchases or spying on user data, and mentioned their trust in universities. Others criticized that the sound was unpleasant. This is a justified criticism, as the sonification has not been designed for a spirit level but for a navigation task, focusing on the highest possible density of unambiguous information, not on learnability or pleasantness. It has not been adopted to the spirit level use case at all, which is practically a dealbreaker in sonification practice \cite[p. 219]{kramer}. 
Unfortunately, many misconceptions and misunderstandings were observable: Some people did not understand that the sonification was the novelty, so they criticized that spirit level apps are nothing new. Others understood that the sound output was the novelty, but did not understand that our sonification contained all the information that the graphics communicated, only with a higher precision. One person correctly commented that using the pitch of a pure tone cannot inform about an absolute position along two axes. Unfortunately, said person was convinced that Tiltification was doing exactly this, based on their interpretation of the teaser video. Some readers correctly understood that one benefit of using sonification was that one could use it whenever it is not possible to view the display, e.g., when crawling under a table to adjust the leg length while the spirit level is placed on the table. Tiltification sparked fruitful debate to whether sending the sensor data to a smartwatch display could be an alternative to using sound, while others thought about binaural audio. These are inspiring thoughts and we are glad that the article about our app has provided ground for constructive thoughts about the topic. In two forums and on YouTube readers commented that the app was useless, as smartphones have an uneven back due to the protruding camera or an affixed phone grip --- which would cause an offset of several degrees. Unfortunately, they did not understand that this is only an issue for graphical solutions, not auditory ones, as the sonification still works when the phone is put on its \textbf{plane} display. In fact, this argument is a great argument \emph{against} all conventional spirit level apps and \emph{for} a sonification based spirit level.

In retrospective, it would have helped to participate actively in these forums to engage with interested people, to break down prejudices and avoid misunderstandings.

\textbf{E-Mail Inquiries:} Around $20$ inquiries reached us via E-Mail and LinkedIn. These contained 
\begin{itemize}
    \item congratulations
    \item suggestions for additional app features
    \item the wish for explanation of the sonification
    \item the wish for an apk, open source and an F-Droid release
    \item complaints that the user survey was only provided in English and not German
\end{itemize}
We also received one extortionate offer to delete negative reviews and add good reviews instead, for a compensation. Some contacted us from their private mail address, others had an institutional signature.

Overall, many users approached us because they were not able to interpret the sonification. It would have been beneficial to add a tutorial with audio examples either inside the app, on the website, or in a YouTube video.

\section{Conclusion}
In this paper we presented our strategy to popularize sonification in general and our sonification app in particular. We reflected on the appropriateness of our measures by analyzing the statistics from the app stores and the responses of media, readers and app users. Despite some mistakes we have made, the app was very successful in terms of the number of downloads and the valence of ratings and reviews. 

Our experience allowed us to give recommendations to those who want to bring their own sonification to the market.


\section{Recommendations}
In the literature you can find so many sonification app prototypes that never made it to the market \cite{steven,alarmclock,musicalutility,araround,blindminton,avissar,SonicTaiji}. We hope that our recommendations help to bring these and other sonification apps to the market.

Even though our two apps may not be representative for sonification app releases in general, we derived several do's and don'ts for the development, distribution and marketing of sonification apps that may serve as an orientation for your own apps:
\subsection{Development}
Developing an app takes much time, effort and expertise, so it is wise to focus on essential aspects. However, make sure to optimize your sonification for your specific application rather than simply implementing an existing sonification.
\subsubsection{Slim and Easy to Use}
For active user engagement, keep your app simple and similar to conventional apps. Explain the sonification interactively in the app, or link to an explanation video. A small file size is appreciated.
\subsubsection{Accessibility}
Make your app accessible to visually impaired people and communicate that.
\subsubsection{Multi-platform and multi-device}
Develop your sonification app cross platform, i.e., for both iOS and Android. Debug and optimize primarily for smartphone usage, rather than tablets. Make sure to have Android and iOS smartphones at hand for testing and debugging.

\subsubsection{Multiple Languages}
Do not waste time and effort on translations: Provide your app and further information in English and your native language.





\subsection{Distribution}
The Apple App Store and Google Play Store are the mainstream channels that you should use. You should automate the deployment and publishing of the apps as described in \cite{autodeploy} and plan at least two working days for the review process.

In addition: Provide an APK under a trustful domain and use a download counter.



\subsection{Marketing}
A university press release is the weapon of choice to advertise your app and raise interest in the media:
\begin{enumerate}
    \item Make sure to address a large audience and not a narrow group
    \item Mention the aspect of accessibility
    \item Prepare citations and interesting photos of involved people
    \item Mention other sonification apps to promote them, too
    \item Provide a project e-mail address and links to the project website and the app store listings
    \item Make sure that you have instructed staff ready to
    \begin{enumerate}
    \item correspond with the press
    \item answer inquiries
    \item engage with readers and users in forums
    \item reply to review in the App stores
    \end{enumerate}
\end{enumerate}

Announce your app release in scientific mailing lists, simultaneously to the press release, to maximize the snowball effect \cite[p. 400]{marketing}. We've made good experience with \href{https://icad.org/mailing-list/}{ICAD},  \href{https://groups.google.com/g/sysmus-conferences}{SysMus}, \href{http://www.auditory.org}{AUDITORY}, \href{https://groups.google.com/g/cec-conference}{cec-c}, \href{https://lists.concordia.ca/cgi-bin/wa?SUBED1=ELECTROACOUSTICS&A=1}{elecroacoustics} and \href{https://www.jiscmail.ac.uk/cgi-bin/webadmin?A0=MUSICOLOGY-ALL}{MUSICOLOGY-ALL}.  
Conference talks and papers are helpful to exchange with the scientific community, but they do not significantly raise the number of downloads.

Do not waste time on creating social media accounts for a short-term project. Instead, focus on essentials:
\begin{enumerate}
    \item Provide a teaser video for the Google Play Store on YouTube
    \item Provide some tutorial videos on YouTube that you can refer to
    \item Engage with partner institutions who have influential social media accounts
\end{enumerate}

Prepare an informative project website under your university sub-domain to
\begin{enumerate}
    \item provide reliable information to the press and to the public
    \item provide photos, graphics and contact info that the press can use
    \item link to the listing in the app stores
    \item host the APK and count the number of APK downloads
    \item generate trust that your app has no financial interest, does not spy on people and that the APK is not harmful.
\end{enumerate}
Either embed YouTube videos with a GDPR consent or host a duplicate of the teaser video to enable people to get informed and download the app while avoiding Google.

\section{Acknowledgment}
\label{sec:ack}
We thank the students of the CURAT bachelor's and CURAT master's project as well as the Sonification Apps master's project, who did an amazing job conceptualizing, designing, implementing, testing, releasing, advertising and maintaining the CURAT Sonification Game and the Tiltification app. We thank Holger Schultheis who was mainly responsible for CURAT. We thank Kevin Austin who forwarded our CURAT advertisement to the cec-conference mailing list and the electroacoustic music mailing list, creating a clear download peak.
\bibliographystyle{IEEEtran}
\bibliography{Bremen}
%
%
%
%

\end{sloppy}
\end{document}